\documentclass[prd,floatfix]{revtex4}

\def\be{\begin{equation}}
\def\ee{\end{equation}}
\def\la{\label}
\def\bea{\begin{eqnarray}}
\def\eea{\end{eqnarray}}
\def\non{\nonumber}
\def\ci{\cite}
\def\la{\label}
\def\bib{\bibitem}

\def\lm{\lambda}
\def\lu{\lambda_1}
\def\ld{\lambda_2}
\def\lt{\lambda_3}

\def\le{\left}
\def\ri{\right}

\def\vp{\varphi}
\def\Omp{\Omega_\phi}

\def\Om{\Omega}

\def\rp{\rho_\phi}

\def\wp{w_\phi}

\def\s8{\sigma_8}

\def\fr{\frac}
\def\pp{\partial}
\def\pu{\pp_\mu}
\def\pU{\pp^\mu}

\def\raw{\rightarrow}

\def\non{\nonumber}

\def\Omp{\Omega_\phi}

\def\r{\rho}
\def\rvp{\rho_\vp}
\def\wvp{w_\vp}
\def\Omvp{\Omega_\vp}
\def\rp{\rho_\phi}

\def\rb{\rho_b}
\def\wb{w_b}

\def\Ob{\Omega_b}

\def\wpe{w_{\phi eff}}
\def\wvpe{w_{\vp eff}}

\def\gb{\gamma_b}

\def\wp{w_\phi}

\def\we{w_{eff}}

\usepackage{graphicx}

\begin{document}

\title{Interacting dark energy: generic cosmological evolution for two scalar fields}

\author{A. de la Macorra }
\affiliation{Instituto de F\'{\i}sica,\\
Universidad Nacional Aut\'onoma de M\'exico,
\\ Apdo. Postal 20-364,
01000 D.F.  M\'exico }

\begin{abstract}

We study the cosmological evolution of two coupled scalar fields
with an arbitrary interaction term $V_T(\phi,\vp)$ in  the presence of
a barotropic fluid,
which can be matter or radiation. The force between the
barotropic fluid and the scalar fields is only
gravitational. We show that the dynamics is completely
determine by only three parameters $\lm_i,\,i=1,2,3$.
We determine   all critical points and
 study their stability. We find six different attractor
 solutions depending on the values of $\lm_i$
 and  we calculate the relevant
 cosmological parameters.
We discuss the possibility of having
one of the scalar fields as   of dark energy
while the other could be a scalar field redshifting as matter.

\end{abstract}


\maketitle

\section{Introduction}

A dark energy component is probably responsible for the present stage
of acceleration of our universe\ci{DE},\ci{SN}.
Perhaps the most appealing candidate for  dark
energy is that of a scalar field, quintessence \ci{Q}, which can be
either a fundamental particle or a composite particle \ci{Qax}.
Within the context of field theory and particle physics
 it is appealing  to interpret  the dark energy as some kind
 of particles  that interact with the particles of the standard model
 very weakly \ci{ax.rp}.
 The weakness of the interaction is required since dark
 energy particles have not been produced in the accelerator
and because the dark energy has not decayed into lighter (e.g. massless) fields
such as the photon.   It is common
to assume the interaction between the dark energy and all other
particles to be via gravity only, however recently interacting dark
energy models have been proposed \ci{IDE}-\ci{wapp}.
This has been motivated in part by the cosmological observations where
an equation of state of dark energy may  be
smaller than minus \ci{DE},\ci{SN}.
In general fluids with $w<-1$ give many
theoretically problems such as stability issues or wrong kinetic
terms as   phantom fields \ci{ph.etc}.  However  interacting dark
energy \ci{IDE}-\ci{wapp},\ci{fate}, where the
dark energy interacts not only gravitationally with other fluids,
is a very simple and
attractive  option which may lead an apparent equation
of state smaller than -1 \ci{wapp}. These fluids can be  dark matter,
neutrinos or other scalar fields.

We study in this letter  the cosmological evolution of the
two scalar fields with arbitrary potentials in the presence of a barotropic
fluid, which can be matter or radiation.
We show that all models dependence lies on three parameters
$\lm_i, i=1,2,3$ defined in eq.(\ref{lm}). We determine the dynamical
equations and obtain the attractor solutions as a function
of these $\lm_i$. We find six different
attractor solutions depending on the relative size of
$\lm_i$ and we calculate the relevant cosmological parameters.

This {\it letter} is organized as follows. In section \ref{sf}
we set up the framework for the cosmological evolution
of two scalar fields with an arbitrary potential in the presence
of a barotropic fluid. In section \ref{de} we present
the conditions on the scalar potential for dark energy.
In section \ref{gda} we derive the dynamical first order
differential  equations and we show that the system
is determined by only three parameters. In section \ref{cs}
we calculate the critical points and we study the stability
of each solution and we give a few examples. In section \ref{lim}
we study different asymptotic limits and we present
a discussion on specific particle physics motivated examples. Finally in section
\ref{con} we present our conclusions.

\section{Coupled Scalar Fields}\la{sf}

Our starting point is a universe  filled with two
scalar fields $\phi,\vp$ and a barotropic  energy density $\rb$,
which can be either matter $\wb=0$ or radiation $\wb=1/3$.
We will assume that the  scalar fields interact
via a potential $V_T(\phi,\vp)$ while there is only gravitational
interaction between these fields and the barotropic fluid. This work generalizes that
of  a single scalar field and a barotropic fluid \ci{miogen}.

One of this
scalar fields, namely $\phi$, may be considered as dark energy
(quintessence) but it is not necessary to  interpret
$\phi$ as dark energy and we will work in a completely general
framework. We take the following Lagrangian for the scalar fields $\phi$ and
$\vp$
\be\la{LT}
L=\fr{1}{2}\pu \phi\pU\phi+\fr{1}{2}\pu\varphi\pU\varphi -
V_T(\phi,\vp)
\ee
with the total potential $V_T(\phi,\vp)$. Instead
of working with the total potential $V_T$ we find
it useful to separate the contribution of the  interacting
and non interacting terms
to differentiate  the contribution of the two
scalar fields. So,  without loss of generality,
we take the total potential as
\be\la{VT}
V_T(\phi,\vp)=V(\phi)+B(\phi,\vp)
\ee
with $V(\phi)$ a potential only
depending on $\phi$ and the interacting potential $B(\phi,\vp)$ is
a function of both fields. Of course we could take
$B(\phi,\vp)=h(\phi,\vp)+K(\vp)$ with $V_T=V(\phi)+h(\phi,\vp)+K(\vp)$
to have a more symmetric potential between the fields $\phi$ and $\vp$. However,
it is more convenient to keep only
the two potentials $V$ and $B$.

The equation of motion of $\phi$ and
$\vp$ for a spatially flat Friedman--Robertson--Walker  (FRW)
universe are
\bea\la{dp}
\ddot\phi+3H\dot\phi&=&-V_{T\phi}= -V_\phi-B_\phi\\
\ddot\vp+3H\dot\vp&=&-V_{T\vp}=-B_\vp
\la{dvp}\eea
where  the subindex in $V$ and $B$ is defined as $V_\phi\equiv \pp V/\pp\phi,
 B_\phi\equiv\pp B/\pp\phi$ and
$B_\vp\equiv\pp B/\pp\vp$.
The Hubble parameter $H\equiv\dot a/a$ is
\be\la{H}
3H^2=\rho =\rp+\rvp+\rb
\ee
where we have taken $8\pi G\equiv 1$ and $\rho$ is the total energy density,
$\rb$   the barotropic fluid and $\rp,\rvp$ are defined in eqs.(\ref{rp})
and (\ref{rvp}).
The mass of the scalar fields is given by
\bea\la{m}
m^2&\equiv&\fr{\pp^2V_T}{\pp\phi^2}= V_{\phi\phi}+B_{\phi\phi}\\
M^2&\equiv&\fr{\pp^2V_T}{\pp\vp^2}= B_{\vp\vp}.
\la{M}\eea
We can also define the energy density and
pressure for the fields $\phi $   as
\be\la{rp}
\rp \equiv\fr{1}{2}\dot\phi^2+  V(\phi),\hspace{.5cm} p_\phi \equiv\fr{1}{2}\dot\phi^2 -
V(\phi)
\ee
and that of $\vp$ as
\be\la{rvp}
\r_\vp \equiv\fr{1}{2}\dot\vp^2+  B(\phi,\varphi), \hspace{.5cm} p_\vp\equiv\fr{1}{2}\dot\vp^2 -
B(\phi,\varphi).
\ee
Using eqs.(\ref{rp}) and (\ref{rvp}) we can rewrite the dynamical
eqs.(\ref{dp}), (\ref{dvp}) in terms of the energy densities as
\bea\la{dr}
\dot\rp+3H\rp(1+\wp) &=&-\delta = -B_\phi\,\dot\phi\non\\
\dot\rvp+3H\rvp(1+\wvp) &=&\delta= B_\phi\,\dot\phi\\
\dot\rb+3H\rb(1+w_b) &=&0\non
\eea
where we have  included the evolution of the barotropic fluid $\rb$ and
\be\la{B}
\delta\equiv B_\phi\,\dot\phi
\ee
defines the interaction term.
The equation of state parameters are given by
 \be\la{wt}
\wp\equiv\fr{p_\phi}{\rp}= \fr{\fr{1}{2}\dot\phi^2-V}{\fr{1}{2}\dot\phi^2+V},\hspace{1cm}
 \wvp\equiv\fr{p_\vp }{\rvp}= \fr{\fr{1}{2}\dot\vp^2-B}{\fr{1}{2}\dot\vp^2+B}
\ee
and the time derivative of $H$ is
\be\la{dH}
\dot H =-\fr{1}{2}\le(\rp+\rvp+\rb+p_\phi+p_\vp+p_b \ri)=-\fr{1}{2}\le(\dot\phi^2+\dot\vp^2+\rb(1+w_b)  \ri).
\ee

\subsection{Effective   Equation of State}\la{ees}

To obtain an effective equation of state we simply
rewrite eqs.(\ref{dr})  as
\be\la{rw}
\dot\rp = -3H\rp(1+w_{eff} ), \hspace{1cm}
\dot\rb = -3H\rb(1+w_{b eff})
\ee
with the effective equation of state defined by \la{fate}
\be\la{weff}
w_{\phi eff} \equiv  \wp + \fr{B_{\phi}\dot\phi}{3H\rp}, \hspace{.5cm}
w_{\varphi eff} \equiv w_b- \fr{B_{\phi}\dot\phi}{3H\r_\vp}.
\ee
We see from eqs.(\ref{rw}) that $\wpe,\wvpe$ give the complete evolution of
$\rp$ and $\rvp$.
For $B_{\phi}\dot\phi>0$ we have $w_{ eff}>\wp$ and
the fluid $\rp$ will dilute faster then without the interaction
term  (i.e. $B_{\phi}\dot\phi=0$) while $\rvp$ will dilute slower since $\wvpe>\wvp$.
Which fluid dominates at late time will depend on which
effective equation of state is smaller. The difference in
eqs.(\ref{weff}) is \ci{miogen}
\be\la{dw}
\Delta w_{eff}\equiv \wvpe-\wpe=\Delta w - \Upsilon
\ee
with $\Delta w\equiv \wvp-\wp$ and  $\Upsilon$ defined as
\be\la{Us}
\Upsilon=  \fr{B_{\phi}\dot\phi}{3H} \le( \fr{\rp+\r_\vp}{\rp\r_\vp}\ri)
\ee
while the sum gives
\be\la{sw}
\Omvp \wvpe+\Omp\wpe= \Omvp \wvp+\Omp\wp.
\ee
Clearly the relevant quantity to determine the relative growth
is given by $\Upsilon$
and if   $\Upsilon > \Delta w$
 we have  $\Delta w_{eff}<0$ and $\rvp$ will dominate the universe at late times
 while for $\Upsilon < \Delta w$ we have  $\Delta w_{eff}>0$ and
 $\rp$ will prevail. In the limit of no interaction $\delta=B_{\phi}\dot\phi=0$
 we get  $\Upsilon=0$ and
$\Delta \we=\Delta w>0$ if $\wp<\wvp$ and $\rp$ will dominate  at late times.
If $\Upsilon = \Delta w$ then $\wvpe=\wpe$ and
 the ratio of both fluids $\rvp/\rp$ will approach a constant value.
If the universe is dominated by
$\rp+\rvp$, i.e. $\Omp+\Omvp=1$, then eq.(\ref{sw}) gives
\be
\wp\leq\; \wpe= \wvp\Omvp+\wp(1-\Omvp) \;\leq \wvp,
\ee
i.e. the effective equation of state is constraint between $\wp$ and $\wvp$.

\subsection{Effective Potential $V_T$}

The effective potential $V_T$ is defined in eq.(\ref{VT}) and
we expect the fields $\phi, \vp$ to evolve to the minimum of the potential,
i.e. $V_{T\phi}=V_\phi+B_\phi \rightarrow 0$ and $V_{T\vp}=B_\vp \rightarrow 0$.
If we want to interpret $V$ as the dark energy potential,
no fine tuning of the potential $V$ requires that the minimum is at
vanishing potential (i.e. $V=0$) and $V_\phi<0$ (see section \ref{de}),
however including the interaction
term $B$ we can now have $V_{T\phi}=0$ for a non vanishing total potential
$V_T$.  A
minimum of $V_T$ can be reached if
$ B_{\phi} > 0$   since $V_\phi<0$ by hypothesis. Taking $V_{T\phi}=0$ and the time
derivative $\dot V_{T\phi}=V_{T\phi\phi}\dot\phi+V_{T\phi\vp}\dot\vp=
m^2\dot\phi+B_{\phi_\vp}\dot\vp= 0$, where we used eq.(\ref{m}),
 one obtains \ci{wapp}
\be
\dot\phi=-\fr{B_{\phi\vp}\dot\vp}{m^2}\propto a^{-3}
\ee
with the solution of $\ddot\phi+3H\dot\phi\approx 0$ giving \ci{de.fer}
\be
 \dot\phi\propto a^{-3}.
 \ee
While the derivative of the effective potential $V_{T\phi}$
is negative the field $\phi$   evolves to larger
values  and even
in the limit  $V_{T\phi}=0$  we have  with $\dot\phi >0$
and a  positive  the interaction term  $\delta =B_\phi\dot\phi$.
The mass $m$ given by eq.(\ref{m}) becomes \ci{wapp}
\be\la{mm}
m^2=  V_{\phi\phi}+B_{\phi\phi}= B_{\phi\phi}\le(1+\widetilde{\Gamma}\fr{B_{\phi}^2}{BB_{\phi\phi}}
\fr{B}{V}\ri) \simeq  B_{\phi\phi}\le(1+ \widetilde{\Gamma}\fr{  B_{\phi}^2}{B B_{\phi\phi}}
\fr{\Omvp}{\Omp}\ri)
\ee
with $\widetilde{\Gamma}\equiv V_{\phi\phi}V/V_{\phi}^2$ ($\widetilde{\Gamma}_m \gtrsim 1$ if the field
$\phi$ is tracking \ci{Q}) and we have approximated
$\rp \simeq V$ and $\rvp\simeq B$ in then last
equality of eq.(\ref{mm}), valid if the kinetic energy
is small compared to the potential energy.

\section{Dark Energy}\la{de}

We may consider   the scalar field $\phi$ as
the quintessence field, i.e.
 dark energy. In the absence of any interaction
 term with $\vp$, the lagrangian is simply given by
$  L(\phi)=\fr{1}{2}\;\dot\phi^2-V(\phi)$ and
  $V$ would be the potential responsible for
present day acceleration. In this case
 the slow roll constrains
\be\la{sr}
|\fr{V_\phi}{V}| \ll 1, \hspace{1cm} \fr{V_{\phi\phi}}{V}\ll 1
\ee
must  be satisfied at present time.
As a result of the dynamics, the scalar field will evolve to its
minimum  and if we do not wish to introduce any kind of unnatural
constant or fine tuning problem, the minimum of the potential must
have zero energy, i.e. $V|_{min}=V_\phi|_{min}=0$ at $\phi_{min}$ \ci{miogen}.

In the absence of a interaction $B$,   a finite value
$\phi_{min}$ implies that the scalar field $\phi$  oscillates around its
vacuum expectation value (v.e.v.). If the scalar field has a non
zero  mass  or if the potential $V$  admits a Taylor expansion
around $\phi_{min}$ then, using the H$\hat { o}$pital rule, one
has  lim$_{t \rightarrow \infty} |V_\phi/V| =\infty$ and
 the energy density $\rp$ redshifts with
$\wp=(n-2)/(n+2)$, i.e. $\wp=0,1/3$ for $n=2,4$ \ci{miogen}.
On the other hand, if $\phi_{min}=\infty$ then $\phi$
will not oscillate and  $|V_\phi/V|$ will approach either zero,  a finite
constant  or  infinity. Only in the case $|V_\phi/V|$
going to zero or a constant smaller than $\sqrt{2}$ will
the universe accelerate at late times \ci{miogen}.

The absence of an arbitrary
dark energy scale requires the
slow roll conditions to be satisfied also in the future
and therefore   $V$ is a runaway potential
and tends to zero at $\phi\rightarrow \phi_{min}=\infty$.
From eqs.(\ref{sr}) we see that $V_\phi$ and $V_{\phi\phi}$ also
approach zero at late times and
$V_\phi$ is therefore negative. Inflation
occurs in general for $\phi \geq 1$ with a mass $m \simeq H$.

\section{Generic Dynamical Analysis}\la{gda}

To determine the attractor solutions of the differential equations
given in eqs.(\ref{dp}) and (\ref{dvp}) or (\ref{dr}) it is useful to make the following
change of variables \ci{miogen},\ci{copeland}
\bea\la{xy}
x_1 \equiv  \fr{\dot \phi }{ \sqrt{ 6} H},\hspace{1cm} y_1 \equiv  \fr{1}{H}\sqrt{ \fr{V }{   3} }\\
x_2 \equiv  \fr{\dot \vp }{ \sqrt{ 6} H},\hspace{1cm} y_2 \equiv
\fr{1}{H} \sqrt{ \fr{B }{  3} }
\eea
and eqs.(\ref{dr})
and (\ref{dH})  become a set of dynamical differential  equations of first order
\bea \la{cosmo1}
x_{1N}&=& -\le(3+\fr{H_N}{H}\ri) x_1 + \sqrt{3 \over 2}\le( \lu   y_1^2   +   \lt \,  y_2^2 \ri)\non  \\
x_{2N}&=& -\le(3+\fr{H_N}{H}\ri) x_2 + \sqrt{3 \over 2} \;\ld \,  y_2^2 \non      \\
y_{1N}&=& -  \fr{H_N}{H}  y_1- \sqrt{3 \over 2}\; \lu \, x_1 \, y_1 \\
y_{2N}&=& -  \fr{H_N}{H}  y_2- \sqrt{3 \over 2}\le( \lt \, x_1+\ld\, x_2\ri)y_2\non \\
\fr{H_N}{H}&=& -{3 \over 2} \le( 2x_1^2+2x_2^2+ \Ob\gb    \ri) \non
\eea
where
$N$ is the logarithm of the scale factor $a$, $N \equiv ln (a)$, $ \gb \equiv 1+w_b$,
$f_N\equiv df/dN$ for $f=x_i,y_i,H\;(i=1,2)$, $\Ob=1-x_1^2-x_2^2-y_1^2-y_2^2$ and
\be\la{lm}
\lu(N) \equiv - \fr{V_{\phi} }{ V},\;\;\;\ld (N) \equiv - \fr{B_\vp }{ B},\;\;\;\lt (N) \equiv - \fr{B_{\phi} }{ B}.
\ee
Notice that all model dependence in eqs.(\ref{cosmo1}) is through
the three quantities $\lambda_i (N), i=1,2,3$ and the constant parameter $\gb=1+\wb$.
The last eq. of (\ref{cosmo1}) is constraint between $-3 \leq H_N/H \leq 0$ for all values of $x_i,y_i$ and $\gb$,
it takes the value $-3$ when the universe is dominated by the kinetic energy
$x_1^2+x_2^2=1$ while it becomes $H_N/H=0$ when the universe is dominate by a
constant potential $y_1^2+y_2^2=1$.
The set of equations given in eqs.(\ref{cosmo1}) give the evolution of two scalar
fields $\phi,\vp$ with arbitrary potentials
in the presence of a barotropic (perfect) fluid with equation of state $w_b=1-\gb$.
As mentioned in section \ref{sf} the choice of dividing the total potential $V_T$
into $V_T=V(\phi)+B(\phi,\vp)$ is without loss of generality and  it is
convenient in order to distinguish the contribution from
both scalar fields.   If we do not want to
consider the contribution from the  barotropic fluid we can easily
 take the limit
$\gb=0$ in eqs.(\ref{cosmo1})   since all contribution form $\rb$ is given in $H_N/H$ via
the term $\Ob\gb$. For $\Ob\neq 0$ we will assume a barotropic fluid with $0<\gb<2$ and $\gb=1$ for
matter while $\gb=4/3$ for radiation.

We do not assume any equation of state for the scalar fields. This is indeed necessary
since one cannot fix the equation of state and the potential
independently.
 For arbitrary potentials the equation of state for
the scalar fields  $\wp=p_{\phi}/\rho_{\phi},\,\wvp=p_{\vp}/\rvp$ is
determined once
\bea
\Omp=\fr{\rho_\phi}{3H^2}&=&x_1^2+y_1^2,\hspace{1cm} \fr{p_\phi}{3H^2}= x_1^2-y_1^2\\
\Omvp=\fr{\rho_\vp}{3H^2}&=& x_2^2+y_2^2,\hspace{1cm} \fr{p_\vp}{3H^2}= x_2^2-y_2^2
 \eea
have been obtained.
Alternatively we can solve for $x_i, y_i $  using eqs.(\ref{cosmo1}) and
the quantities
$\wp\equiv w_1=(x_1^2-y_1^2)/(x_1^2+y_1^2)$ and $\wvp\equiv w_2=(x_2^2-y_2^2)/(x_2^2+y_2^2)$ are,
in general, time or scale dependent. In terms
of $x_i,y_i$ the interaction term  in eq.(\ref{B}) becomes
\be\la{B1}
\delta=B_{\phi}\dot\phi= -\sqrt{6}\,3H^3 \lt   x_1 y_2^2
\ee
giving an  effective equation of state parameters
defined  in eqs.(\ref{weff}) as
\bea\la{weff2}
\wpe  &=&   \wp -\sqrt{\fr{2}{3}}\;  \fr{\lt\,x_1y_2^2}{\Omp}=
\fr{x_1^2 -y_1^2-\sqrt{\fr{2}{3}}\;  \lt\,x_1y_2^2}{\Omp}\\
w_{\vp eff} &=&  w_\vp+\sqrt{\fr{2}{3}}\;   \fr{\lt \, x_1y_2^2}{\Omvp}
=\fr{x_2^2-y_2^2(1-\sqrt{\fr{2}{3}}\;   \lt \, x_1)}{\Omvp}
\eea
and the acceleration
of the universe is given by
\bea\la{ac}
\fr{\ddot a}{a} &=& H^2+ HH_N = -\fr{H^2}{2}\le(\Ob(1+3\wb)+4(x_1^2+x_2^2)-2(y_1^2+y_2^2)  \ri )\non\\
&=&-\fr{H^2}{2}\le(4-3\Ob(1-\wb) -
  6(y_1^2 + y_2^2)\ri)
\eea
where we have used $\dot H=H H_N$ and eq.(\ref{cosmo1}).
Clearly acceleration will occur if the universe is dominated by
the potential $y_T^2\equiv(y_1^2+y_2^2)=V_T/3H^2=(V+B)/3H^2$, i.e.
 for   $y_T^2>2/3-\Ob(1-\wb)/2$.

\section{Critical Solutions}\la{cs}

We find the critical solutions to   the dynamical equations (\ref{cosmo1})
with $x_{1N}=x_{2N}=y_{1N}=y_{2N}=0$ and solve for constant
values of $\lm_i, i=1,2,3$. The set of solutions are given in tables
\ref{tUxy} and \ref{txy}.  In table \ref{tUxy} we give, for completeness,  the unstable
critical points and   in table \ref{tUw} we show the values of
the equation of state parameters for these solutions. More interesting, we give
in table \ref{txy} the attractor (stable) solutions. Constant $\lm$ implies
that the potential $V$ and $B$ are exponential potentials, e.g.
$V\sim e^{-\alpha \phi}$ with  $\lm_1=-V_{\phi}/V=\alpha$ constant. However, if
the potential is not exponential we do not expect to have constant
$\lm_i$ and they will, in general, evolve with time. In this case we can use the attractor solution
of table \ref{txy} and take the corresponding limit of $\lm_i$, which
will be either zero, constant or infinity \ci{miogen}, to obtain
the asymptotic behavior of the solutions.

In order to determine the stability of the critical points
we perturb eqs.(\ref{cosmo1}) around  the critical solution $x_i,y_i$
and we keep linear terms only . The set of eqs.
 can be written in a matrix form, $Z_N=M Z$, where
 $Z=(\delta x_1,\delta x_2, \delta y_1, \delta y_2)$
 and we diagonalize the matrix $M$.
The stability of the solution requires
the real part of all eigenvalues to be negative.
One of the eigenvalues of  model U-I
is positive with  $Ei=3\gb/2>0$ for $\gb>0$ while all other models in table \ref{tUxy}
have at least one eigenvalue
of the form $Ei=3(2-g)>0$ for $0<g < 2$. Therefore they
are all unstable solutions.

The eigenvalues of models in table \ref{txy}  are given
in table \ref{tEig}. We see that depending on the
values of $\lm_i$ the eigenvalues can be negative
or positive. For any given choice of $\lm_i$ there is
only one stable solution.
In table \ref{tcon} we give the constrains
on $\lm's$ form closure $|x_i|\leq 1, y_i \leq 1$ and stability
arguments. We see that we have two main conditions on $\lm_i$.
One is the relative size between $\lu^2$ and $\lu\lt$ while the
second is the relative size between $  \ld^2+\lt^2$ and $\lu\lt $.
Depending on the relative size of these two conditions the attractor
solution will end up either in models S-I,S-II or S-III,S-IV or S-V, S-VI.
A further condition on the magnitude  of $\lm_i$ and $3\gb$ distinguishes
between the different models.

In tables \ref{tOm} and \ref{tw} we give the values of the relevant
cosmological parameters such $\Omp,\Omvp,\Ob, y^2_T=y_1^2+y_2$ and $\wp,\wvp,\wpe,\wvpe,
g = \sqrt{ 2/3}\lt x_1y_2^2 $. As mentioned in section \ref{ees} the
 effective equation of state gives the correct  redhsift of the scalar field
when the interaction term is considered.
 Attractor solutions S-I and S-II
are the critical solution of a single scalar field in the presence of
a barotropic fluid. In these models the energy density  $\rvp$ redshifts faster than
$\rp$ and $\rb$ so $\wvpe$ must be
larger than $\wpe$ or $w_b$. We see   from table \ref{tw} that models S-III to S-VI
have both scalar fields with the same redshift, i.e. $\wpe=\wvpe$. In models S-III and S-V the redshift is equal to
that of the barotropic fluid while in model S-IV and S-VI it depends on the value of
the different $\lm_i$ with $\wpe=\wvpe<w_b$. A universe dominated by the barotropic fluid is possible
in  models S-I, S-III and S-V, while models with
no barotropic fluid ($\Ob=0$) are given by  S-II, S-IV and S-VI.
An accelerating universe requires
$y_T^2$ to be larger than $2/3-\Ob/2$ and therefore  models S-II, S-IV and
S-VI are favored. From table \ref{tOm} and conditions in table  \ref{tcon}  we can see
that models S-I, S-III and S-V have $y^2_T<1-\gb/2<1/2$  for $\gb\geq 1$ and
therefore do not lead to an accelerating universe.

The eigenvalues of model S-V are
\bea\la{ei5}
Ei_{V,1/2} &= & \fr{3(\gb-2)}{4}\pm \fr{3}{4}\sqrt{\fr{(2-\gb)(24\gb^2[(\lu-\lt)^2+\ld^2]+(2-9\gb)\lu^2\ld^2)}{\lu^2\ld^2}}\\
Ei_{V,3/4} &= &\fr{3(\gb-2)}{4} \pm \fr{3}{4}\sqrt{\fr{ (2-\gb)(8\gb\lt[(\lu-\lt)^2+\ld^2]+(2-9\gb)\lu\ld^2)}{\lu\ld^2}}\non
\eea
while of model S-VI are
\bea\la{ei6}
Ei_{VI,1} &= & \fr{\lu^2\ld^2-3\gb[(\lu-\lt)^2 + \ld^2] }{(\lu-\lt)^2 + \ld^2} \non\\
Ei_{VI,2} &= &\fr{  \lu^2\ld^2 -6[(\lu-\lt)^2 + \ld^2]}{2[(\lu-\lt)^2 + \ld^2] }\\
Ei_{VI, 3/4} &= & \fr{  \lu^2\ld^2 -6[(\lu-\lt)^2 + \ld^2]}{4[(\lu-\lt)^2 + \ld^2] }\pm\fr{1}{4} \fr{\sqrt{a_{VI}}}{((\lu-\lt)^2 + \ld^2)}
\non
\eea
with
\bea
a_{VI}&\equiv &  2ABC+2D^2-BE-F^2\non\\
A&\equiv &(\lu-\lt)^2 + \ld^2,\hspace{1cm} B\equiv \lu^2\ld^2-6A,\hspace{1cm} C\equiv3\gb-\lu\lt,\\
D&\equiv & 3A(2+\gb)-2\lu^2\ld^2, \hspace{.5cm} E \equiv A(2\lu\ld-6(1+2\gb))+3\lu^2\ld^2,\hspace{.5cm}F\equiv6A(1-\gb)+\lu^2\ld^2\non
\eea
The effective equation of state are given by
 \bea\la{we}
 (\wpe) &=& \wp + \fr{g}{\Omp},   \hspace{1cm}
 (\wvpe) = \wvp - \fr{g}{\Omvp}
 \eea
with $g$ given in table \ref{tw}.
For model S-VI we have
 \bea\la{w6}
 (\wp)_{VI}&=&  \fr{\lu^2\ld^4-  (6[(\lu-\lt)^2 + \ld^2]-\lu^2\ld^2)(\ld^2+\lt^2-\lu\lt) }
 {\lu^2\ld^4+  (6[(\lu-\lt)^2 + \ld^2]-\lu^2\ld^2)(\ld^2+\lt^2-\lu\lt)  }\non \\
 (\wvp)_{VI}&=&  \fr{ \lu  \ld^2(2\lu  -  \lt )   - 6 [(\lu-\lt)^2 + \ld^2 ]}
 {(6[(\lu-\lt)^2 + \ld^2] - \lu \ld^2  \lt  )}.
 \eea

\subsection{Examples}

We now present four  different attractor solution depending on the values of $\lu,\ld,\lt$.
We show in figures \ref{fig1}-\ref{fig4}  the evolution of $\Omp\equiv \Om_1, \Omvp\equiv\Om_2,\Ob$
 and  $ w_1\equiv\wp,\wpe,  w_2\equiv\wvp ,  \wvpe$ for the different choices of $\lm's$.  We also show
 the phase space of  $(x_1,y_1)$ and $(x_2,y_2)$ for each case. Since the phase space depends
 on four variables, namely  $(x_1,y_1,x_2,y_2)$ , it is no surprising that the curves in
 the two dimensional space $(x_1,y_1)$ and $(x_2,y_2)$ may cross.

In figs.\ref{fig1}
we have $\lu^2=5,\ld=1,\lt=3$ and $\gb=1+\wb=1$ which implies that conditions
of model S-I in table \ref{tcon} are satisfied, i.e. $3=3\gb<\lu^2=5$ and $5=\lu^2<\lu\lt\simeq 6.7$.
In this case  $\rp$ and $\rb$ have the same redshift at late times and $\Omp/\Ob$ approaches
a constant value while $\rvp$ redshifts faster
with an effective $\wvpe>w_b=0$, even though $w_2=\wvp\rightarrow -1$, and therefore $\Omvp$ tends to zero.
The attractor solution has $(x_1,y_1)=(\sqrt{3/10},\sqrt{3/10})\simeq (0.55,0.55)$ and $(x_2,y_2)=(0,0)$
and $\Om_1= 0.6$, $\Om_2= 0$ with  $\Ob=0.4$ and $w_b=\wp=\wvpe=0$ while $\wvp=-1$ and $\wvpe=0.34$.
Since $y_T^2=y_2^2+y_2^2=0.3$   is smaller than $2/3-\Ob/2=0.87$ then from eq.(\ref{ac}) we
conclude that  there is no late time acceleration.

In figs.\ref{fig3}
we take $\lu=2,\ld=1/2,\lt=1/2$ and $\gb=1+\wb=1$ which implies that conditions
of model S-IV in table \ref{tcon}  are met, i.e. $0.5=\ld^2+\lt^2<\lu\lt=1$ and $3=3\gb>\ld^2+\lt^2=0.5$.
In this case  $\rp$ and $\rvp$  have the same redshift at late times and $\Omp/\Omvp$ approaches
a constant value while $\Ob$ goes to zero. The
effective equation of state $\wpe$ and $\wvpe$ have the same late time value  $\wpe=\wvpe\simeq -0.83$ while
$\wp=1$ and $\wvp=-0.91$.
The attractor solution has $(x_1,y_1)=(0.2,0)$ and $(x_2,y_2)=(0.2,0.96)$
and $\Om_1= 0.04$, $\Om_2= 0.96$ and $\Ob=0$.
In this case $y_T^2=y_2^2+y_2^2=0.92$    is larger than $2/3-\Ob/2=2/3$ and  from eq.(\ref{ac}) we
conclude that the universe   accelerates at late times.

In figs.\ref{fig2}
we have  $\lu=3,\ld=2,\lt=3/2$ and $\gb=1+\wb=1$ which implies that conditions
of model S-V in table \ref{tcon}  are satisfied, i.e. $6.25=\ld^2+\lt^2>\lu\lt=4.5$, $9=\lu^2>\lu\lt=4.5$,
$18.75=3\gb[(\lu-\lt)^2+\ld^2]<\lu^2\ld^2=36$  and $9=\lu^2>3\gb^2/2=3/2$.
In this case  $\rp, \rvp$ and $\rb$  have all the same redshift at late times and $\Omp,\Omvp, \Ob$ approach
a constant value. The
effective equation of state $\wpe$ and $\wvpe$ have the same late time value  $\wpe=\wvpe=\wb\simeq 0$ while
$\wp=0.4$ and $\wvp=-1/3$.
The attractor solution has $(x_1,y_1)=(0.4,0.27)$ and $(x_2,y_2)=(0.3,0.43)$
and $\Om_1=\Omp=0.24$, $\Om_2=\Omvp=0.29$ and $\Ob=0.47$.
Since $y_T^2=y_2^2+y_2^2=0.26$   is smaller than $2/3-\Ob/2\simeq 0.43$ then from eq.(\ref{ac}) we
conclude that   the universe does not   accelerate at late times.

Finally, in figs.\ref{fig4}
we take  $\lu=1,\ld=1,\lt= -3$ and $\gb=1+\wb=1$ which implies that conditions
of model S-VI in table \ref{tcon} are satisfied, i.e. $10=\ld^2+\lt^2>\lu\lt=-3$, $1=\lu^2>\lu\lt=-3$,
$51=3\gb[(\lu-\lt)^2+\ld^2]>\lu^2\ld^2=1$  and $17\sqrt{6}=\sqrt{6}|\lu|[(\lu-\lt)^2+\ld^2]>\lu^2\ld^2=1$.
In this case  $\rp$ and $\rvp$  have the same redshift at late times and $\Omp/\Omvp$ approaches
a constant value while $\Ob$ goes to zero. The
effective equation of state $\wpe$ and $\wvpe$ have the same late time value  $\wpe=\wvpe\simeq -0.98$ while
$\wp=-0.99$ and $\wvp=-0.92$. Notice that at some redshifts the effective equation of state
take values $\wpe<-1$.
The attractor solution has $(x_1,y_1)=(0.02,0.87)$ and $(x_2,y_2)=(0.1,0.48)$
and $\Om_1=\Omp=0.76$, $\Om_2=\Omvp=0.24$ and $\Ob=0$.
Since $y_T^2=y_2^2+y_2^2=0.99$   is smaller than $2/3-\Ob/2=2/3$ then from eq.(\ref{ac}) we
conclude that   the universe does  accelerate at late times.

\begin{table}
\begin{center}
\begin{tabular}{|c|c|c|c|c|c|c|c|}
  \hline
Model &   $ y_1  $ & $   y_2  $ & $   x_1  $ & $   x_2  $ & $  \Om_1=\Omp $  & $  \Om_2=\Omvp  $ & $   \Ob  $  \\ \hline
\hline
U-I    & $  0 $ & $    0 $ & $    0 $ & $    0 $ & $ 0  $ & $  0 $ & $ 1  $  \\
U-II    & $  0 $ & $    0 $ & $    \pm 1    $ & $    0 $ & $  1 $ & $ 0  $ & $ 0  $ \\
U-III   & $  0 $ & $    0 $ & $    \pm \sqrt{1 - x_2^2}   $ & $   x_2  $ & $  1-x_2^2 $ & $ x_2^2  $ & $ 0 $  \\
U-IV   & $  0 $ & $    0 $ & $    \fr{\sqrt{6}}{\lu}    $ & $    \pm \sqrt{1  -\fr{6 } {\lu}}   $
& $  \fr{6}{\lu^2}  $ & $ 1- \fr{6}{\lu^2}  $ & $ 0  $  \\
U-V    & $  0 $ & $    0 $ & $    \fr{\sqrt{6} \lt  + \ld \sqrt{ \ld^2
+ \lt^2 -6 }}{  \ld^2 + \lt^2}  $ & $
  \fr{\sqrt{6} \ld - \lt\sqrt{\ld^2 + \lt^2 -6}}{\ld^2 + \lt^2}    $
  & $ x_1^2 $ & $  x_2^2 $ & $  0 $  \\
U-VI    & $  0 $ & $    0 $ & $    \fr{\sqrt{6} \lt  - \ld \sqrt{ \ld^2
+ \lt^2 -6}}{  \ld^2 + \lt^2}  $ &
$      \fr{\sqrt{6} \ld + \lt\sqrt{ \ld^2 + \lt^2 -6}}{\ld^2 + \lt^2}     $
& $ x_1^2  $ & $ x_2^2  $ & $  0 $ \\
\hline
\end{tabular}
\end{center}
\caption{\small {Unstable critical solutions. }}
\la{tUxy}\end{table}

\begin{table}
\begin{center}
\begin{tabular}{|c|c|c|c|c|c|  }
  \hline
Model  & $
 \wp  $ & $   \wvp  $  &  $ \wpe  $ & $  \wvpe $ & $   \Upsilon  $  \\ \hline
\hline
U-I   & $ -  $ & $ -  $  & $ -  $ & $ -  $ & $  - $  \\
U-II&   $ 1  $ & $  - $   & $  1 $ & $ -  $ & $ -  $  \\
U-III&   $  1 $ & $ 1  $ & $ 1  $ & $  1 $ & $ 0  $ \\
U-IV&   $ 1  $ & $  1 $& $  1 $ & $  1 $ & $  0 $   \\
U-V &   $  1 $ & $ 1  $ & $ 1  $ & $  1 $ & $ 0  $  \\
U-VI &  $ 1  $ & $  1 $ & $  1 $ & $  1 $ & $ 0  $   \\
\hline
\end{tabular}
\end{center}
\caption{\small {Equation of state parameters of the unstable solutions of table \ref{tUxy}. }}
\la{tUw}\end{table}

\begin{table}
\begin{center}
\begin{tabular}{|c|c|c|c|c|  }
  \hline
Model &   $ y_1  $ & $   y_2  $ & $   x_1  $ & $   x_2  $   \\ \hline
\hline
S-I   & $   \sqrt{ \fr{3 (2 - \gb) \gb }{  2\lu^2 }}    $ & $0$ & $\sqrt{ \fr{3 }{2}}\fr{\gb}{ \lu }     $  & 0 \\
S-II &   $  \sqrt{1 - \fr{\lu^2}{6}}  $ & $    0 $ & $   \fr{\lu}{\sqrt{6}}    $ & $    0 $  \\
S-III &   $  0 $ & $    \sqrt{\fr{3(2 - \gb) \gb}{2(\ld^2 + \lt^2)}} $ & $
\sqrt{\fr{3}{2}} \fr{  \gb \lt}{\ld^2 + \lt^2}    $ &
$    \sqrt{\fr{3}{2}}\frac{ \gb \ld}{ \ld^2 + \lt^2 }    $ \\
S-IV &   $  0 $ & $    \sqrt{1-\fr{   \ld^2+\lt^2 }{ 6} }    $ & $
\fr{\lt}{\sqrt{6}}    $ &$\fr{\ld}{\sqrt{6}} $ \\
S-V    & $ \sqrt{ \fr{ 3( 2 - \gb) \gb (\ld^2 + \lt^2 -\lu \lt)}{2 \lu^2 \ld^2 }}    $  &
$   \sqrt{ \fr{3 (2 - \gb) \gb (\lu - \lt)}{2\lu \ld^2}}  $ &
$    \sqrt{\fr{3}{2}}\fr{ \gb}{\lu}  $ & $     \sqrt{\fr{3}{2}}\fr{ \gb (\lu - \lt)}{\lu \ld}   $ \\
S-VI   &  $\fr{\sqrt{ (6[(\lu-\lt)^2 + \ld^2]-\lu^2\ld^2))(\ld^2+\lt^2-\lu\lt)    }}{\sqrt{6}((\lu-\lt)^2 + \ld^2) } $
& $ \fr{\sqrt{  \lu (\lt - \lu) (\lu^2\ld^2  - 6 [(\lu-\lt)^2 +\ld^2] )}}{\sqrt{6}((\lu-\lt)^2 + \ld^2)  } $ &
 $   \fr{\lu \ld^2}{\sqrt{6} ((\lu-\lt)^2 + \ld^2   )}    $ &
 $    \fr{\lu \ld (\lu - \lt)}{\sqrt{6} ((\lu-\lt)^2 + \ld^2   )} $ \\
  \hline
\end{tabular}
\end{center}
\caption{\small {Stable critical solutions.  }}
\la{txy}\end{table}

\begin{table}
\begin{center}
\begin{tabular}{|c|c|c|c|c|c|  }
 \hline
Models&   $ Ei_1  $ & $  Ei_2  $ & $Ei_3  $ & $  Ei_4 $   \\ \hline
\hline
S-I &     $ \fr{3}{2}(\gb-2)    $ & $ \fr{3}{2} \gb(1-\fr{\lt}{\lu})    $ &
$ \fr{3}{4}(\gb-2)  + \fr{3}{4}\sqrt{\fr{(2-\gb)(24\gb^2+(2-9\gb)\lu^2)}{\lu^2}}   $ &
$\fr{3}{4} (\gb-2) - \fr{3}{4}\sqrt{\fr{(2-\gb)(24\gb^2+(2-9\gb)\lu^2)}{\lu^2}} $ \\
S-II    &  $   \fr{1}{2}(6-\lu^2 ) $ & $  \fr{1}{2}(6-\lu^2)   $ &
$ -3\gb+\lu^2  $ & $ \fr{1}{2}\lu(\lu-\lt)  $ \\
S-III&      $\fr{3}{2}(\gb-2)   $ & $ \fr{3\gb(\ld^2+\lt^2-\lu\lt)}{2(\ld^2+\lt^2)}  $ &
$ \fr{3}{4}(\gb-2) +\fr{3}{4}\sqrt{\fr{(2-\gb)(24\gb^2+(2-9\gb)(\ld^2+\lt^2))}{\ld^2+\lt^2 } }  $ &
$ \fr{3}{4}(\gb-2) -\fr{3}{4}\sqrt{\fr{(2-\gb)(24\gb^2+(2-9\gb)(\ld^2+\lt^2))}{ \ld^2+\lt^2  } }  $    \\
S-IV &      $  \fr{1}{2}(\ld^2+\lt^2-6)  $ & $   \fr{1}{2}(\ld^2+\lt^2-6)    $ & $   \ld^2+\lt^2-3\gb   $
 & $   \fr{1}{2} (\ld^2+\lt^2-\lu\lt)  $  \\
S-V &     $  Ei_{V,1} $ & $  Ei_{V,2} $ & $ Ei_{V,3}  $ & $ Ei_{V,4} $   \\
S-VI &   $  Ei_{VI,1} $ & $  Ei_{VI,2} $ & $  Ei_{VI,3} $ & $  Ei_{VI,4} $   \\
  \hline
\end{tabular}
\end{center}
\caption{\small {Eigenvalues for the different stable  solutions of table \ref{txy}. Stability
requires  the real part of all  eigenvalues to be negative. Eigenvalues $Ei_{V}$
are given in eqs.(\ref{ei5}) while  eigenvalues $Ei_{VI}$
 in eqs.(\ref{ei6}).}}
\la{tEig}\end{table}

\begin{table}
\begin{center}
\begin{tabular}{|c|c|c|     }
 \hline
  Model &  $ Constraint$ & $  Constraint $    \\ \hline
\hline
 S-I &   $  \lu^2> 3\gb$ & $  \lu^2<\lu\lt $      \\
  S-II &  $  \lu^2<3\gb $ & $  \lu^2<\lu\lt $    \\
    S-III&   $ \ld^2+\lt^2<\lu\lt  $ & $   \ld^2+\lt^2 >3\gb $       \\
 S-IV &  $   \ld^2+\lt^2<\lu\lt $ & $\ld^2+\lt^2  < 3\gb $      \\
  S-V &    $  \ld^2+\lt^2>\lu\lt $ & $ \lu^2> \lu\lt  $   \\
 &   $ \lu^2> 3\gb^2/2$&   $ 3\gb[(\lu-\lt)^2+\ld^2]<\lu^2\ld^2  $    \\
  S-VI &   $ \ld^2+\lt^2>\lu\lt  $ & $  \lu^2>\lu\lt $     \\
  &    $\sqrt{6}|\lu|[(\lu-\lt)^2+\ld^2]>\lu^2\ld^2$ & $3\gb[(\lu-\lt)^2+\ld^2]>\lu^2\ld^2$ \\
  \hline
\end{tabular}
\end{center}
\caption{\small {Constrains on $\lu,\ld,\lt$ for the different stable solutions of table \ref{txy}
 from closure and stability arguments. }}
\la{tcon}\end{table}

\begin{table}
\begin{center}
\begin{tabular}{|c|c|c|c|c|  }
  \hline
Model &     $  \Om_1=\Omp $  & $  \Om_2=\Omvp  $ & $   \Ob  $   & $ y_T^2=y_1^2+y_2^2   $  \\ \hline
\hline
S-I &  $  \fr{3\gb}{ \lu^2 }   $ & $ 0  $ & $ 1- \fr{3\gb}{ \lu^2 }  $ & $ \fr{3(2-\gb)\gb}{2\lu^2}   $  \\
S-II &   $ 1  $ & $ 0  $ & $  0 $ & $  1-\fr{\lu^2}{6}  $ \\
S-III &    $   \fr{ 3 \gb^2 \lt^2}{2(\ld^2 + \lt^2)^2}   $ &
$  \fr{ 3 \gb(2\ld^2+(2-\gb)\lt^2)  }{2(\ld^2 + \lt^2)^2}   $ &
$ 1-\fr{3\gb}{\ld^2 + \lt^2}   $ & $   \fr{3(2-\gb)\gb}{2(\ld^2+\lt^2)}   $ \\
S-IV    &  $   \fr{\lt^2}{6} $
  & 1-$ \fr{\lt^2}{6}   $ & 0 & $  1-  \fr{\ld^2+\lt^2}{6} $ \\
S-V &     $ \fr{3\gb(2\ld^2-\lt(2-\gb)(\lu-\lt))}{2\lu^2\ld^2}  $ &
$  \fr{3\gb(\lu-\lt)(2\lu-\gb\lt)}{2\lu^2\ld^2}   $ & $ 1- \fr{3\gb((\lu-\lt)^2+\ld^2)}{\lu^2\ld^2}   $
& $ \fr{3(2-\gb)\gb[(\lu-\lt)^2+\ld^2]}{2\lu^2\ld^2}   $ \\
S-VI
 & $ \fr{ (6[(\lu-\lt)^2 + \ld^2]-\lu^2\ld^2))(\ld^2+\lt^2-\lu\lt)+\lu^2\ld^4  }{6[(\lu-\lt)^2 + \ld^2]^2} $ &
 $  \fr{  \lu ( \lu-\lt) (6 [(\lu-\lt)^2 +\ld^2]-\lu\ld^2\lt   )  }
{6[(\lu-\lt)^2 + \ld^2]  } $ & $ 0 $ & $1 -\fr{\lu^2\ld^2}{6[(\lu-\lt)^2+\ld^2]}    $ \\
  \hline
\end{tabular}
\end{center}
\caption{\small {Energy density for the different stable  solutions of table \ref{txy}. }}
\la{tOm}\end{table}

\begin{table}
\begin{center}
\begin{tabular}{|c|c|c|c|c|c|  }
 \hline
Model &   $ \wp  $ & $   \wvp  $ & $ \wpe  $ & $  \wvpe $ & $   g = \sqrt{\fr{2}{3}}\lt x_1y_2^2 $  \\ \hline
\hline
S-I &   $w_b  $ & $  - $ & $w_b   $ & $ > w_b,\wpe  $ & $ 0 $  \\
S-II &   $-1+\fr{\lu^2}{3}   $ & $  - $ & $ -1+\fr{\lu^2}{3}$ & $ > w_b, \wpe   $ & $ 0 $ \\
S-III &    $ 1  $ & $  -1 +\fr{2\gb\ld^2}{2\ld^2+(2-\gb)\lt^2}  $ & $ w_b $ & $ w_b  $
& $ \fr{3(2-\gb)\gb^2\lt^2}{2(\ld^2+\lt^2)}$   \\
S-IV &    $  1 $ & $ -1 +\fr{2\ld^2 }{6-\lt^2}   $ & $ -1 + \fr{\ld^2+\lt^2}{3}$ & $  -1 + \fr{\ld^2+\lt^2}{3}  $
& $ \fr{\lt^2(\ld^2+\lt^2-6)}{18}  $  \\
S-V &     $ -1+ \fr{2\gb\ld^2}{2\ld^2+(2-\gb)(\lt-\lu)\lt} $ &
 $ \fr{2(1-\gb)\lu+\gb\lt}{\gb\lt-2\lu} $ &
 $ w_b $ & $ w_b  $
  & $ \fr{3(2-\gb)\gb^2\lt(\lu-\lt)}{2\lu^2\ld^2}$ \\
S-VI &   $ (\wp)_{VI} $ & $  (\wvp)_{VI} $ & $-1+\fr{\lu^2\ld^2}{ 3[(\lu-\lt)^2 + \ld^2]} $ &
$ -1+\fr{\lu^2\ld^2}{3[ (\lu-\lt)^2 + \ld^2]}  $
&  $ \fr{ \lu^2\ld^2\lt(\lu-\lt)(6[(\lu-\lt)^2 + \ld^2]-\lu^2\ld^2)}{18[(\lu-\lt)^2 + \ld^2]^3}   $ \\
  \hline
\end{tabular}
\end{center}
\caption{\small {Equation of state parameters of the stable
solutions of table \ref{txy}. The equation of state parameters $ w_{VI} $ are given
in eqs.(\ref{w6}).}} \la{tw}\end{table}

\begin{figure}[htp!]
\begin{center}
\includegraphics[width=8cm]{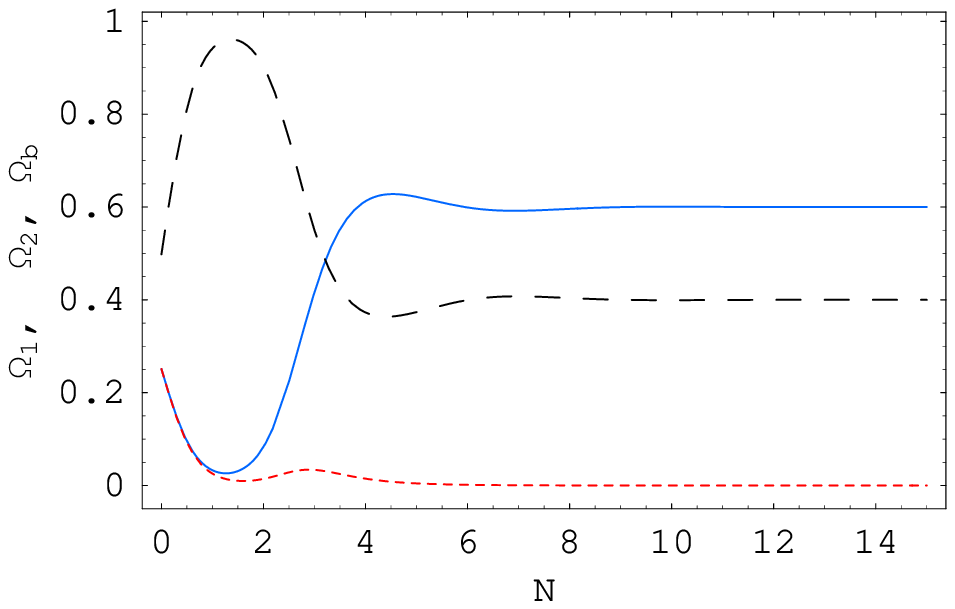}
\includegraphics[width=8cm]{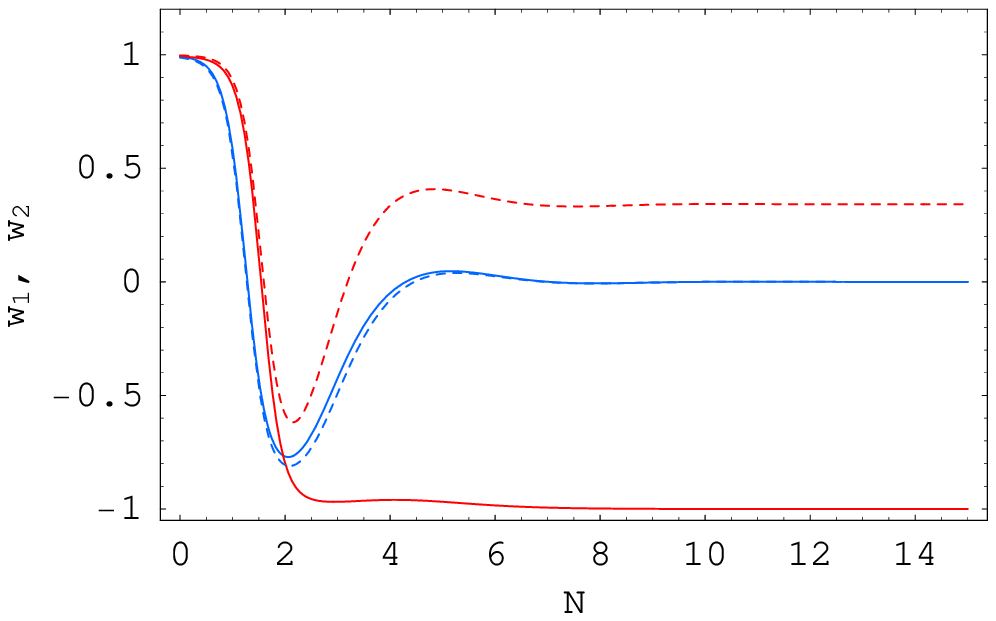}
\includegraphics[width=8cm]{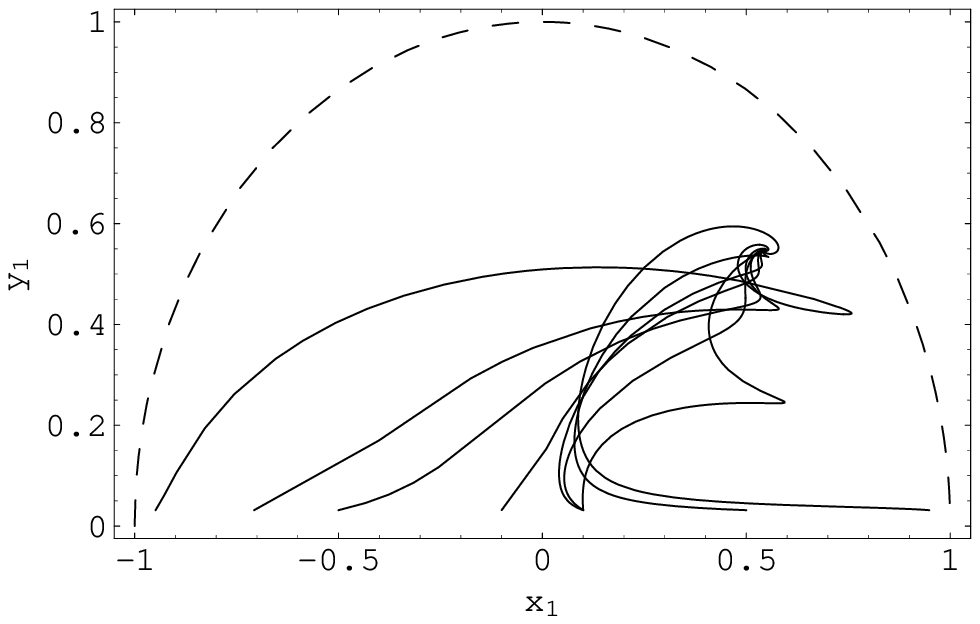}
\includegraphics[width=8cm]{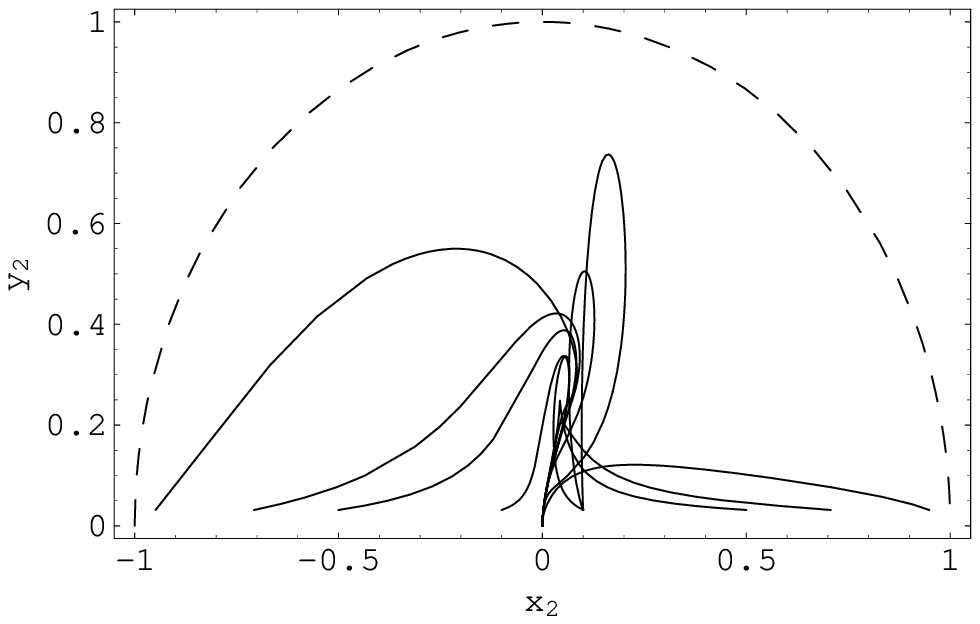}
\end{center}
\caption{\small{We show  the evolution of $\Omp\equiv \Om_1,
\Omvp\equiv\Om_2,\Ob$ (blue (solid), red (dotted)  and black
(dashed), respectively) and the equation of state parameters
$\wp=w_1,\wpe$ (blue (solid), blue (dotted))   and $\wvp=w_2,
\wvpe$  (red (solid), red (dotted)) as a function of $N=Log[a]$,
for $\lu^2=5,\ld=1,\lt=3$ and $\gb=1+\wb=1$. With these choice of
$\lm's$ the attractor solution is given by model S-I. The
attractor solution has $(x_1,y_1)=(\sqrt{3/10},\sqrt{3/10})\simeq
(0.55,0.55)$ and $(x_2,y_2)=(0,0)$. }}
\la{fig1}
\end{figure}

\begin{figure}[htp!]
\begin{center}
\includegraphics[width=8cm]{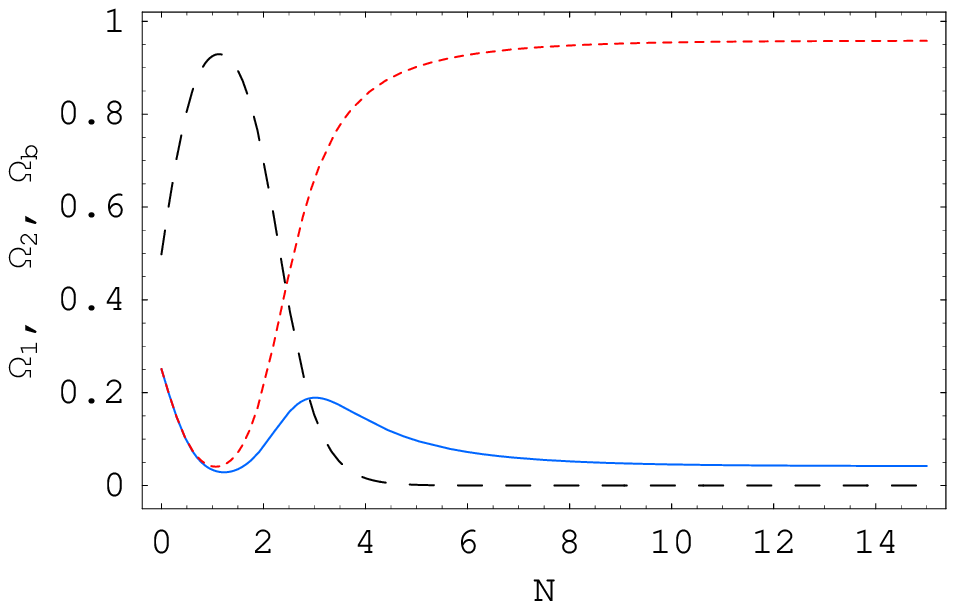}
\includegraphics[width=8cm]{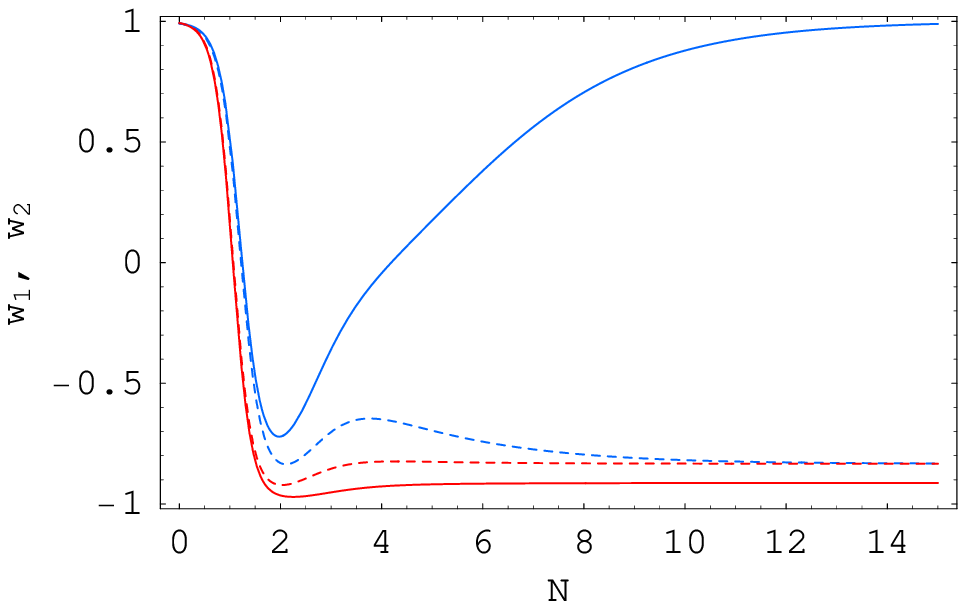}
\includegraphics[width=8cm]{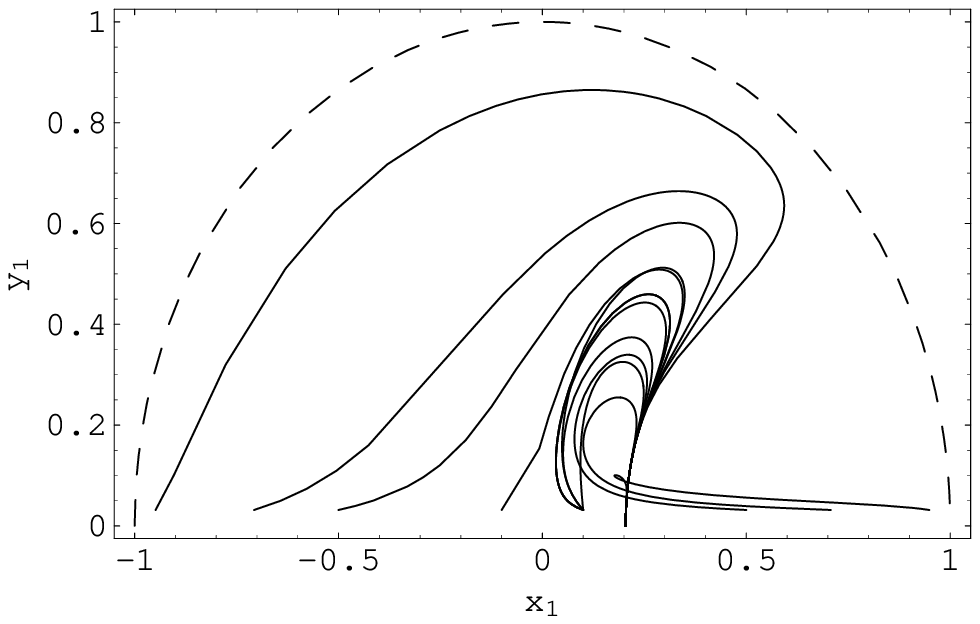}
\includegraphics[width=8cm]{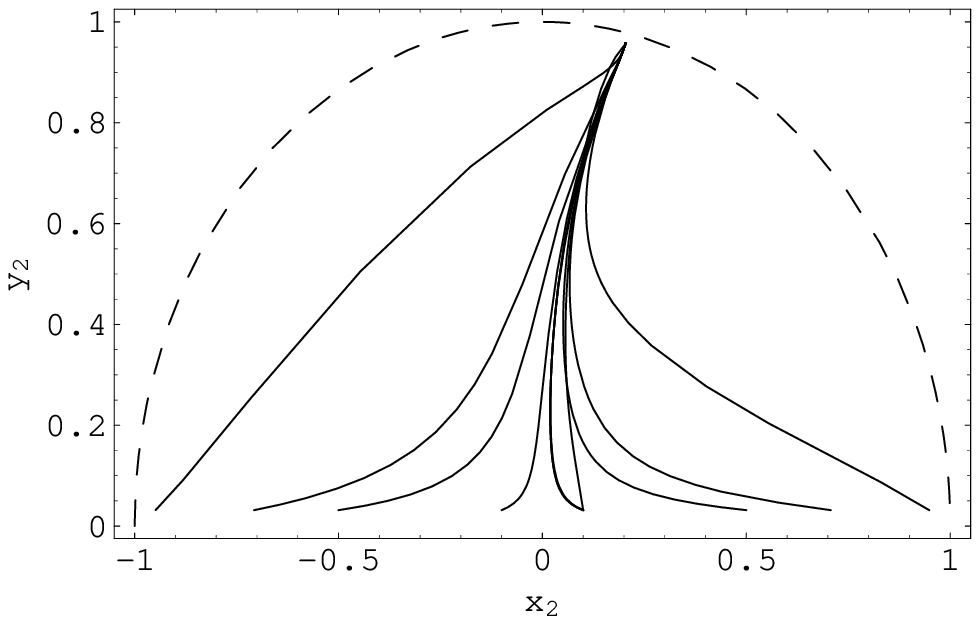}
\end{center}
\caption{\small{We show  the evolution of  $\Omp=\Om_1,
\Omvp=\Om_2, \Ob$ (blue (solid), red (dotted)  and black (dashed),
respectively) and the equation of state parameters $\wp=w_1,\wpe$
(blue (solid), blue (dotted))   and $\wvp=w_2,  \wvpe$  (red
(solid), red (dotted)) as a function of $N=Log[a]$, for
$\lu=2,\ld=1/2,\lt=1/2$ and $\gb=1+\wb=1$. With these choice of
$\lm's$ the attractor solution is given by model S-IV. The
attractor solution has $(x_1,y_1)=(0.2,0)$ and
$(x_2,y_2)=(0.2,0.96)$. }}
\la{fig3}
\end{figure}

\begin{figure}[htp!]
\begin{center}
\includegraphics[width=8cm]{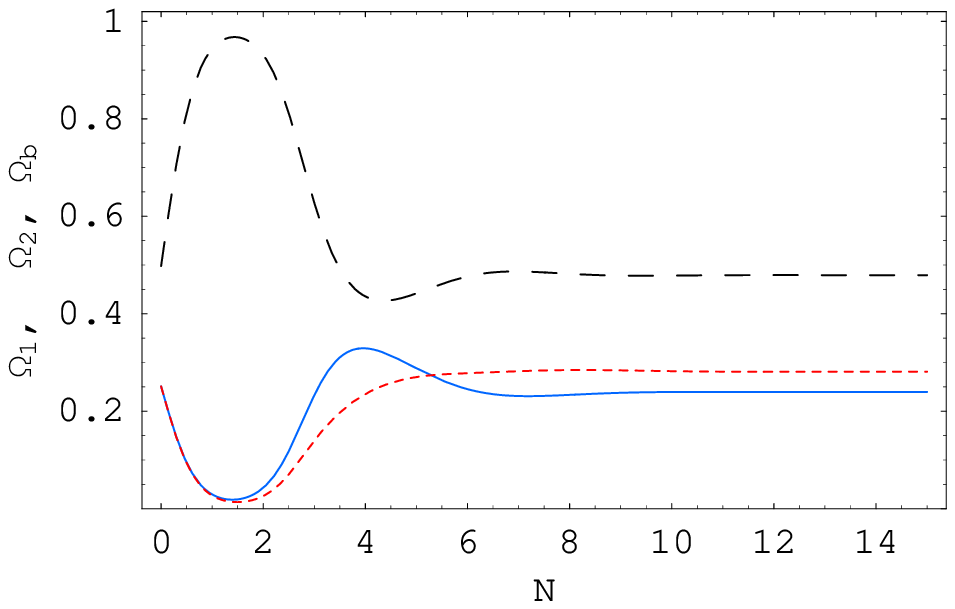}
\includegraphics[width=8cm]{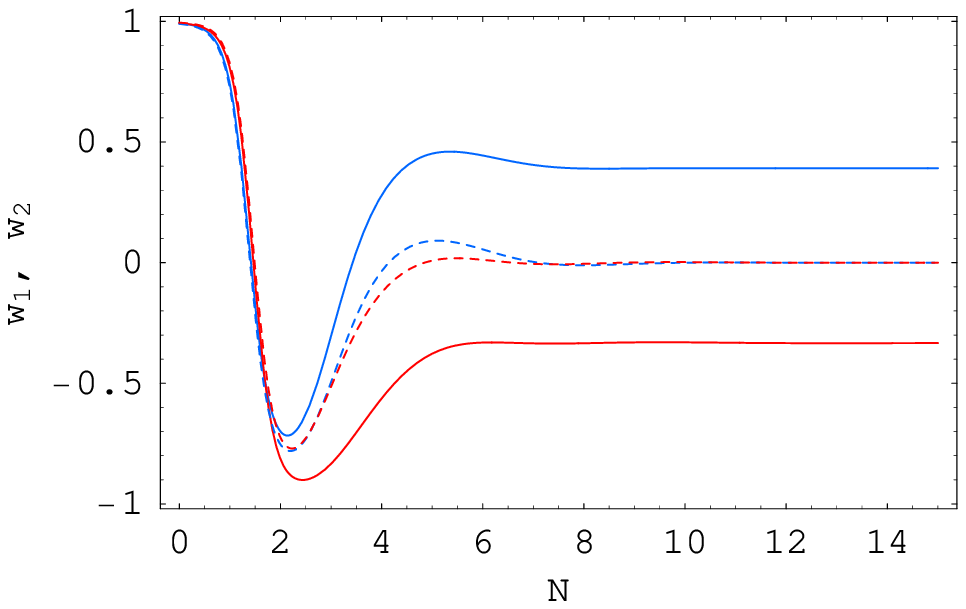}
\includegraphics[width=8cm]{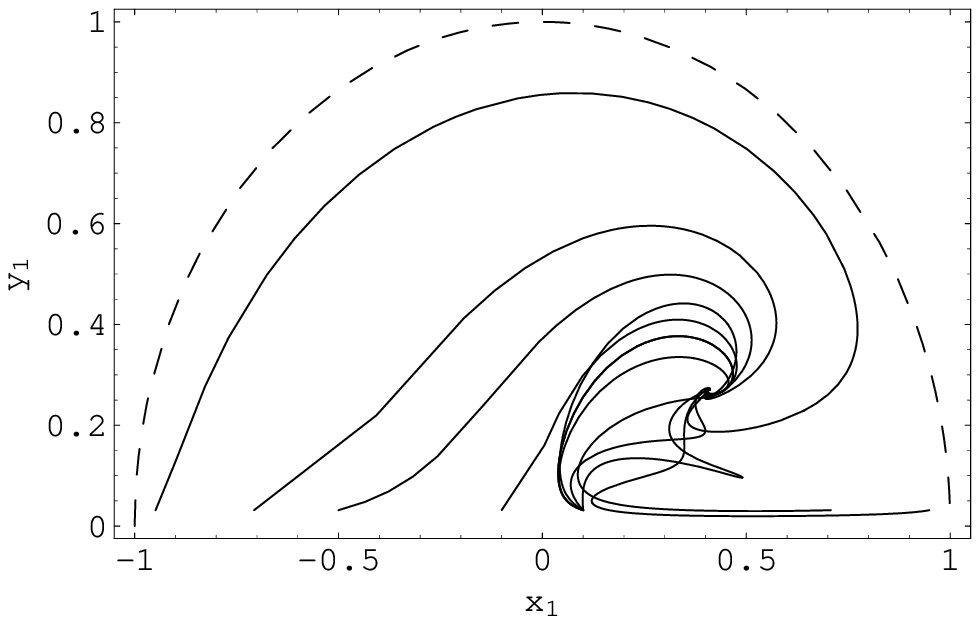}
\includegraphics[width=8cm]{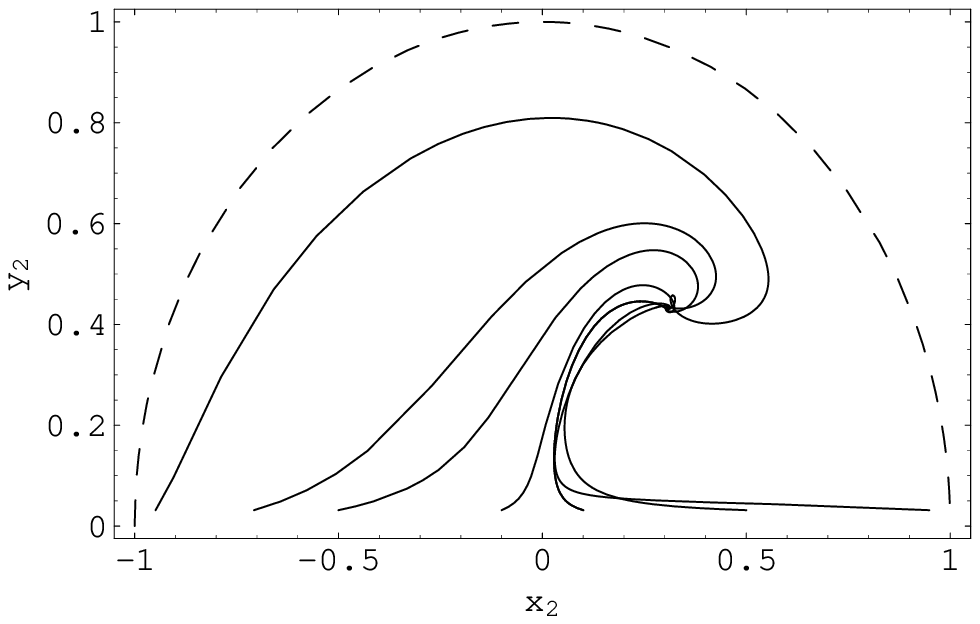}
\end{center}
\caption{\small{We show  the evolution of $\Omp=\Om_1,
\Omvp=\Om_2,\Ob$ (blue (solid), red (dotted)  and black (dashed),
respectively) and the equation of state parameters $\wp=w_1,\wpe$
(blue (solid),  blue (dotted))   and $\wvp=w_2,  \wvpe$  (red
(solid), red (dotted)) as a function of $N=Log[a]$, for
$\lu=3,\ld=2,\lt=3/2$ and $\gb=1+\wb=1$.With these choice of
$\lm's$ the attractor solution is given by model S-V.  The
attractor solution has $(x_1,y_1)=(0.4,0.3)$ and
$(x_2,y_2)=(0.3,0.4)$.  }}
\la{fig2}
\end{figure}

\begin{figure}[htp!]
\begin{center}
\includegraphics[width=8cm]{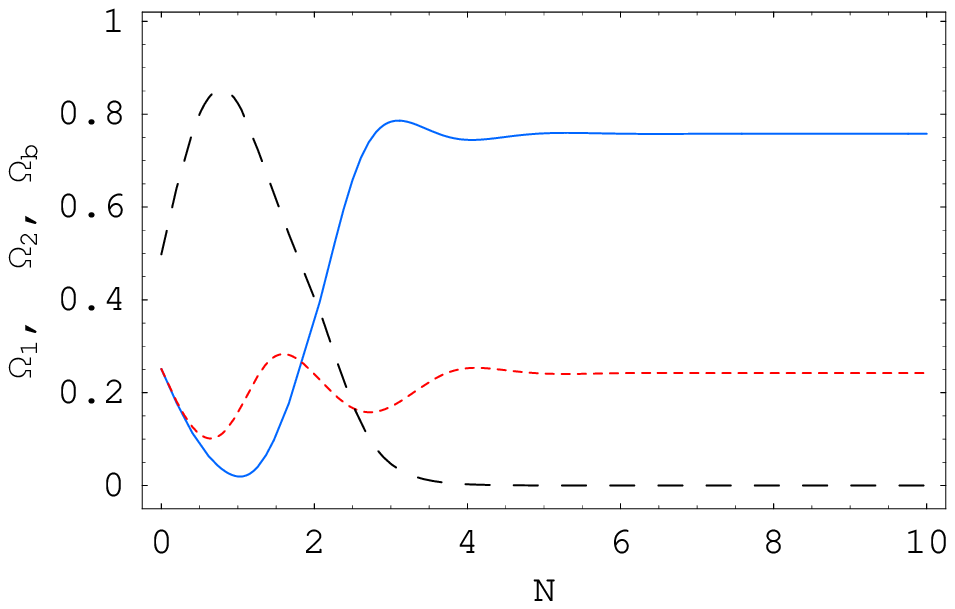}
\includegraphics[width=8cm]{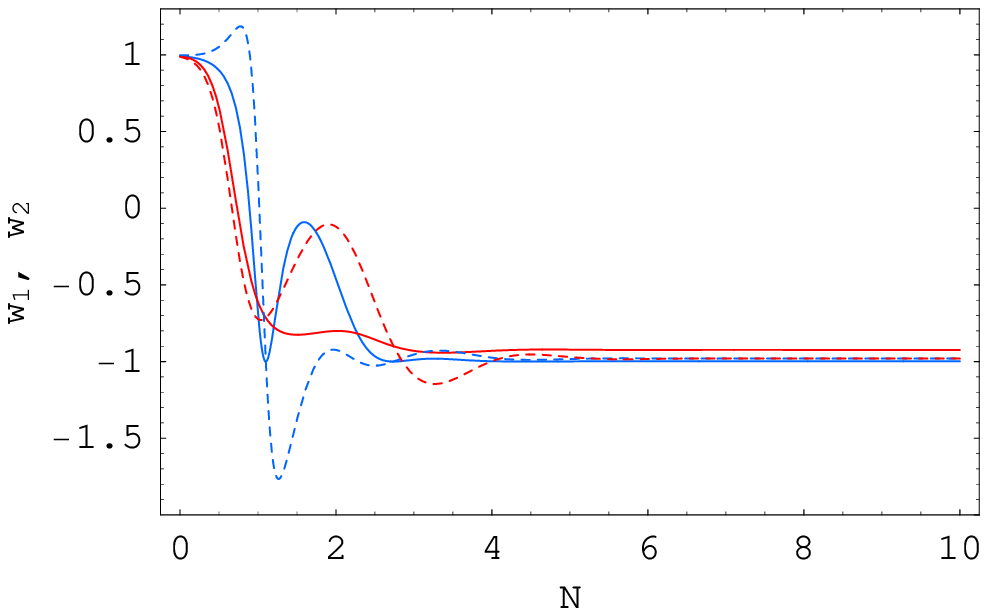}
\includegraphics[width=8cm]{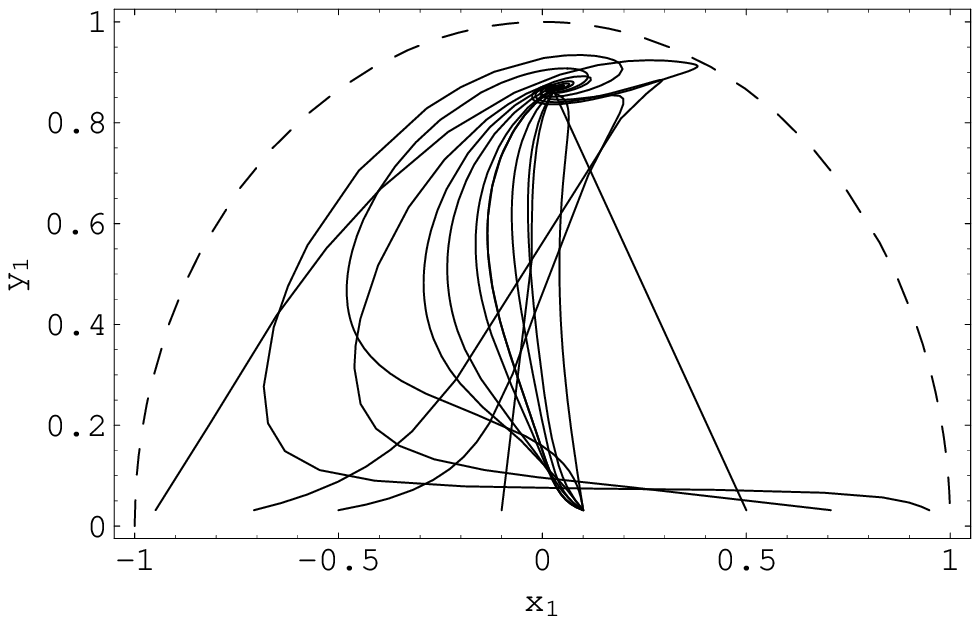}
\includegraphics[width=8cm]{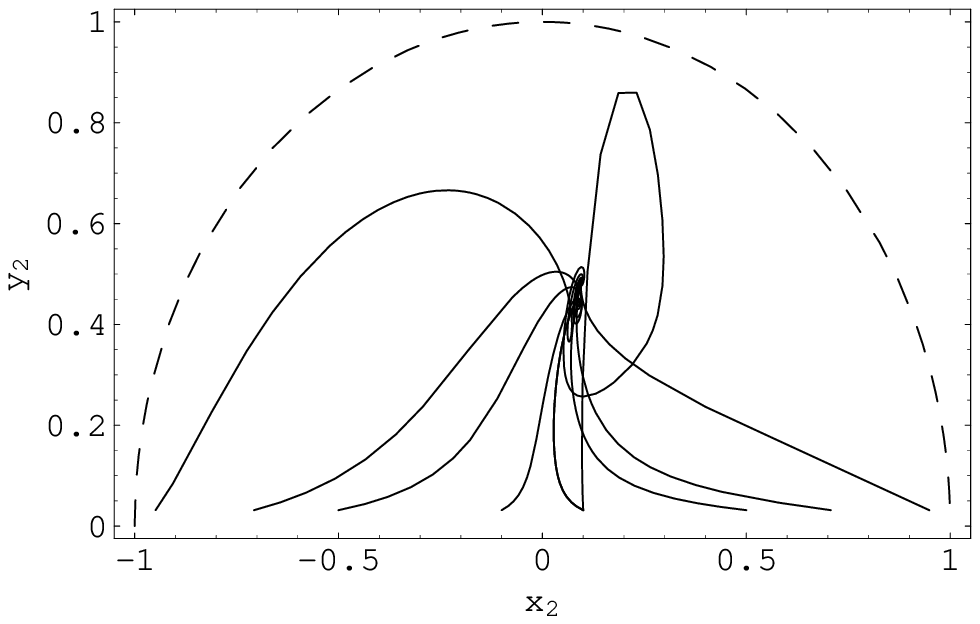}
\end{center}
\caption{\small{We show  the evolution of $\Omp=\Om_1,
\Omvp=\Om_2,\Ob$ (blue (solid), red (dotted)  and black (dashed),
respectively) and the equation of state parameters $\wp=w_1,\wpe$
(blue (solid), blue (dotted))   and $\wvp=w_2,  \wvpe$  (red
(solid), red (dotted)) as a function of $N=Log[a]$, for
$\lu=1,\ld=1,\lt=-3$ and  $\gb=1+\wb=1$. With these choice of
$\lm's$ the attractor solution is given by model S-VI. The attractor solution
has $(x_1,y_1)=(0.02,0.87)$ and $(x_2,y_2)=(0.1,0.52)$}}
\la{fig4}
\end{figure}

\section{Asymptotic Behavior}\la{lim}

Special cases can be studied by taking different limits of the parameters $\lm$.
From tables \ref{tcon} and \ref{tw} we can determine which values of $\lm_i$ are required for
any  particular late time behavior we wish to have. For example if we are interested
in the behavior of having only two scalar fields and no barotropic fluid ($\Ob=0$) we can take
the limit $\gb=0$ of models in table \ref{txy}. Only models S-II, S-IV and S-VI survive
and $x_i,y_i$ take the same values as in  table \ref{txy}. If we prefer a
universe dominated by the barotropic fluid than models S-I, S-III and
S-V need to be considered. An accelerating universe requires
$y_T^2$ to be larger than $2/3-\Ob/2$ and therefore  models S-II, S-IV and
S-VI are favored.

If we take the limit $\lu=0$ then only   models S-II and S-VI remain consistent
and they have $(x_1,y_1)=(0,1)$ and $(x_2,y_2)=(0,0)$.
Models S-I and S-V do not satisfy the closure condition $|x_i|\leq 1|, y_i\leq 1$)
while models S-III and S-IV are no longer stable if $\ld\neq 0\neq \lt$ (c.f. eigenvalues of table \ref{tEig}).
In the limit $\lu \raw 0$, the  first derivative of the potential
approaches zero faster than the potential itself and examples of
this kind of behavior are given by potentials of the form $V=V_0
\phi^{-n}, n >0$.

For $\ld=0$ only model S-V  is no longer consistent.
Models S-I and S-II  remain the same as in table \ref{txy}
and model S-III becomes $(x_1,y_1)=(\sqrt{3/2}\;\gb/\lt,0)$ and $(x_2,y_2)=(0,(\sqrt{3(2-\gb)\gb/2}/\lt)$,
S-IV is now  $(x_1,y_1)=(\lt/\sqrt{6},0)$ and $(x_2,y_2)=(0,(\sqrt{1- \lt^2/6})$ while
S-VI becomes   $(x_1,y_1)=(0,\sqrt{\lt/(\lt-\lu)})$ and $(x_2,y_2)=(0,\sqrt{\lu/(\lu-\lt)})$.

In the case $\lt=0$ all models remain valid from closure arguments
and are given in table \ref{tL30}. However,  models S-I to S-IV
are no longer stable for $\lu\neq 0\neq \ld$, as seen from table \ref{tEig},
and only models S-V and S-VI are the attractor solutions.
\begin{table}
\begin{center}
\begin{tabular}{|c|c|c|c|c|  }
  \hline
Model &  $ y_1  $ & $   y_2  $ & $   x_1  $ & $   x_2  $   \\ \hline
S-I &   $   \sqrt{\fr{3(2 - \gb) \gb)}{2\lu^2 }}$ & $ 0$ & $\sqrt{\fr{3}{2}} \fr{\gb}{\lu}$ & $ 0$  \\
S-II &  $ \sqrt{1 - \fr{\lu^2}{6}}$ & $ 0$ & $ \fr{\lu}{\sqrt{6}}$ & $ 0 $ \\
S-III &     $ 0$ & $   \sqrt{\fr{3 (2 - \gb) \gb)}{2\ld^2}}$ & $ 0$ & $ \sqrt{\fr{3}{2}}\fr{\gb}{\ld} $ \\
S-IV &     $ 0$ & $ \sqrt{1 -\fr{\ld^2}{6}}$ & $ 0$ & $ \fr{\ld}{\sqrt{6}} $ \\
S-V &     $   \sqrt{ \fr{3(2 - \gb) \gb}{2\lu^2}}$ & $   \sqrt{\fr{3(2 - \gb) \gb}{2\ld^2}} $ &
   $ \sqrt{\fr{3}{2}} \fr{\gb}{\lu} $ & $ \sqrt{\fr{3}{2}}\fr{ \gb}{\ld} $ \\
S-VI &     $ \fr{ \sqrt{\ld^2 [6 (\ld^2 + \lu^2) -\lu^2 \ld^2]}}{\sqrt{6}(\lu^2 + \ld^2)}$ & $
  \fr{\sqrt{  \lu^2    (6 [( \lu ^2 +\ld^2]-\lu^2\ld^2  )}}
{\sqrt{6}(\lu ^2 + \ld^2)  }$ & $
    \fr{\lu \ld^2}{\sqrt{6} (\lu^2 + \ld^2)}$ & $ \fr{\lu^2 \ld}{\sqrt{6} (\lu^2 + \ld^2)} $ \\
  \hline
\end{tabular}
\end{center}
\caption{\small {Attractor solutions of table \ref{txy} in the limit $\lt=0$ }}
\la{tL30}\end{table}

Finally, if we impose the condition $\lu y_1^2+\lt y_2^2=0$, which minimize the potential as
a function of $\phi$, i.e $dV_T/d\phi=d(V+B)/d\phi=0$, there are only three attractor solutions
to eqs.(\ref{cosmo1}). The first case is model S-II $(x_1,y_1)=(0,1)$ and $(x_2,y_2)=(0,0)$ (c.f.
table \ref{txy} in the limit $\lu=-\lt y_2^2/y_1^2=0$). The other two models are S-III and S-IV of
table \ref{txy} with  the limit $\lt=-\lu y_1^2/y_2^2=0$ (c.f. table \ref{tL30}).
There is no stable solution with $\lm_i,\,i=1,2,3$ constant and $y_1y_2\neq0$.
Let us take the limit $\lu y_1^2+\lt y_2^2=0$ in eqs.(\ref{cosmo1}).
The evolution for $x_1$   is
\be\la{xy1}
\frac{x_{1N}}{x_1}=
-(3+\frac{H_N}{H})\leq 0, \hspace{2cm}
\la{lm=0}
\ee
and since $-3 \leq H_N/H \leq 0$ for all values of
$x_i,y_i$ and $\gb$ we conclude that $x_1$ will
approach its minimum value (i.e. $x_1\raw 0$).
If $|\lu|<\infty$, so that $\lt x_1\rightarrow 0$, then the evolution of $y_1$ becomes
\be
\frac{y_{1N}}{y_1}= -  \fr{H_N}{H}  - \sqrt{3 \over 2}\; \lu \, x_1\simeq -  \fr{H_N}{H}   \geq 0
\ee
and  $y_1$ will increase
to its maximum value (i.e. $y_1 \raw 1$). Since $x_1^2+y_1^2+x_2^2+y_2^2+\Ob=1$
then  for  $y_1= 1$ we get $x_1=x_2=y_2=\Ob=0$ giving model S-II.

\subsection{Particle Physics Model}

Let us now take a specific choice of  potentials motivated by particle
physics. We consider a
factorisable  interaction potential
\be
 B(\phi,\vp)=h(\phi)F(\vp).
 \ee
In this case the $\lm_i$ parameters
become only functions of a single field
\be
\lu(\phi)=-\fr{V'(\phi) }{V(\phi)},\hspace{.3cm}
\ld(\vp)=-\fr{F'(\vp)} {F(\vp)},\hspace{.3cm}
\lt(\phi)=-\fr{h'(\phi) }{h(\phi)}
\ee
where the prime denotes
derivative w.r.t. its argument. Let us take a simple example
where the potential for the scalar field $\phi$ has an exponential
behavior, widely used in particle physics as for dark energy potential,
while we take a power law potential for $\vp$ which would represent
a standard scalar field. It is well known that
for a  single scalar field with potential of the form $\vp^n$
the energy density redshifts as $\wvp= (n-2)/(n+2)$ giving
$\wvp=0,1/3$ for $n=2,4$ \ci{miogen}. So, we take
$V=V_oe^{-\alpha\phi},
 B=B_o e^{\beta\phi}\vp^n$ with $V_o>0, B_o>0$. In this
 case $\lu=\alpha, \lt=-\beta, \ld=-n/\vp$ . The effective
 potential defined in eq.(\ref{VT})  is minimized for
 \bea
V _{T \phi} &=& V_\phi + B_\phi=-\alpha V +\beta B=0 \\
V _{T \vp } &=&   B_\vp = n B_o e^{-\beta\phi} \vp^{n-1}=0
\eea
which implies  $V/h=(V_o/B_o) e^{-(\alpha+\beta)\phi}=(\beta/\alpha) \vp^n$, i.e.
\be
\phi= -Log[A\vp^n]/(\alpha+\beta), \hspace{.5cm}
h =  e^{-[\beta/(\alpha+\beta)]   Log[A\vp^n] } =
  A^{-\beta/(\alpha+\beta)} \vp^{-n\beta/(\alpha+\beta)}
\ee
with $  A=B_o\beta/V_o\alpha$ and  $A$ should be positive
and we take $\alpha>0, \beta>0$.
 The condition $h\vp^{n-1}\rightarrow 0$
becomes
\be
h\vp^{n-1}=
h_o A^{-\beta/(\alpha+\beta)} \vp^{ [(n-1)\alpha+n\beta]/(\alpha+\beta)}\rightarrow 0
\ee
and for $[(n-1)\alpha+n\beta]/(\alpha+\beta)>0$ (valid for $n>1$) we find $\vp\rightarrow 0$
giving $  |\ld|\rightarrow\infty$. From table
\ref{tcon} we see that for $  |\ld|\gg |\lu|,|\lt| $ the stable
solution are  given by  model
S-V or  S-VI depending whether $3\gb$ is smaller or larger
than $\lu^2=\alpha^2$. Model S-V has
$(\Om_1,\Om_2,\Ob,y_T^2)=(3\gb/\lu^2,0,1-3\gb/\lu^2,3(2-\gb)\gb/2\lu^2)$
and model S-VI has
$(\Om_1,\Om_2,\Ob,y_T^2)=(1,0,0,1-\lu^2/6)$. We see that in both
cases the amount of energy density from the field $\vp$ vanishes
and  the solutions reduce to a single
scalar field $\phi$ which depending on the value of $\lu=\alpha$ we can have
a universe completely dominated by the scalar field
$\phi$ with an accelerating universe for $\lu^2<2$  or a scalar field
redshifting as the barotropic fluid for $\lu^2>3\gb$. The interaction term $g$
given in table \ref{tw} vanishes for models S-V, S-VI in the limit
$|\ld|\rightarrow\infty$. The equation of state for $\vp$
 for model S-V is $\wvp=
(2(1-\gb)\lu+\gb\lt)/(\gb\lt-2\lu) $ while
 for model S-VI we have $\wvp= (6+\lu\lt-2\lu^2)/(\lu\lt-6)$   (c.f. eq.(\ref{w6})).

\section{Summary and Conclusions}\la{con}

We have studied the dynamical system of two
scalar fields with arbitrary potentials in the
presence of a barotropic fluid in a FRW metric. We have shown
that all  model dependence is given in terms
of three parameters, namely $
\lu(N) = - \ V_{\phi}/V,\,
\ld (N) = -  B_\vp /B,\,  \lt(N) = - B_{\phi}/B$,
and we have calculated all critical points
and determine their stability. We have seen
that there are six  different attractor solutions given  in table
\ref{txy}. For a given choice of $\lm_i$ the attractor solution depends
on the relative magnitude of $\lu,\ld,\lt$ and $\gb$
as discussed in table \ref{tcon}. We have calculated   the
relevant cosmological parameters and we have shown
the phase space for four different attractor solutions.
Finally, we have discussed the different asymptotic limits
of the parameters $\lm_i$ and we have studied
a special type of scalar potential motivated by particle physics.

\section*{Acknowledgments}\non

This work was also supported in
part by CONACYT project 45178-F and DGAPA, UNAM project
IN114903-3.

\end{document}